\documentclass[a4paper,amsmath,amssymb]{jpconf}
\usepackage{graphicx}
\usepackage{amsmath, amsthm, amssymb}
\begin{document}
\title{ Spreading of Non-Newtonian and Newtonian Fluids on a Solid Substrate under Pressure}

\author{Moutushi Dutta Choudhury$^1$, Subrata Chandra$^1$, Soma Nag$^1$, Shantanu Das$^2$
and Sujata Tarafdar$^1$
}

\address{$^1$ Condensed Matter Physics Research Centre, Physics Department, Jadavpur University, Kolkata 700032, India\\
$^2$ Reactor Control Division, Bhabha Atomic Research Center, Trombay, Mumbai 400085, India}

\ead{mou15july@gmail.com}

%%%%%%%%%%%%%%%%%%%%%%%%%%%%%%%%%%%%%%%%%%%%%%%%%%%%%%%
%%                                      Abstract                                                                                             %%
%%%%%%%%%%%%%%%%%%%%%%%%%%%%%%%%%%%%%%%%%%%%%%%%%%%%%%%
\begin{abstract}
Strongly non-Newtonian fluids namely, aqueous gels of starch, are shown to exhibit visco-elastic behavior, when subjected to a load. We study arrowroot and potato starch gels. When a droplet of the fluid is sandwiched between two glass plates and compressed, the area of contact between the fluid and plates increases in an oscillatory manner. This is  unlike Newtonian fluids, where the area increases monotonically in a similar situation. The periphery moreover, develops an instability, which looks similar to Saffman Taylor fingers. This is not normally seen under compression. The loading history is also found to affect the manner of spreading. We attempt to describe the non-Newtonian nature of the fluid through  a visco-elastic model incorporating generalized calculus. This is shown to reproduce qualitatively the oscillatory variation in the surface strain.
\end{abstract}
%%%%%%%%%%%%%%%%%%%%%%%%%%%%%%%%%%%%%%%%%%%%%%%%%%%%%%%%%
\section{Introduction}
\label{sec:Introduction}
%%%%%%%%%%%%%%%%%%%%%%%%%%%%%%%%%%%%%%%%%%%%%%%%%%%%%%%%%
 
Static and dynamic aspects of wetting and spreading are well studied classical text book problems. However,
forced spreading of a fluid under an impressed load is not widely discussed,  though the physics involved is challenging and the problem has important applications in technology as well. In real life many fluids are non-Newtonian, adding further complexity to the problem. General reviews concerning related problems are available \cite{Engmann,bonn,nag}, but there is scope for more experimental and analytical studies.

In the present work we report studies on spreading of non-Newtonian fluids between two glass plates, when the upper plate is loaded by a weight. We vary the weights, the fluids and the manner of loading. 

We observe an interesting oscillation in the area of contact between the fluid and glass plate as a function of time. Earlier study of Newtonian fluids \cite{nag,aps} did not show such behavior. Based on the experimental results we try to explain this phenomenon using  fractional calculus, which is known to be an appropriate technique to study non-Newtonian, visco-elastic materials \cite{visco,das1,das2}. Another remarkable observation is the appearance of a surface instability similar to viscous fingering. Viscous fingering under the condition of lifting, i.e. separating the plates is a well studied phenomenon \cite{epje,martine,gay}. In this case the pressure is lower within the fluid, compared to the air pressure outside, satisfying the Saffman-Taylor condition for instability \cite{saffman}. However, in the present case the fingering develops during {\it compression}.\\

\section{Materials and Methods}

\subsection{Materials}
The fluids under study are two non-Newtonian fluids - arrowroot and potato starch gel and a Newtonian fluid ethylene glycol (GR) is also studied for comparison.

In order to get a homogenous gel of the starch in water the suspension has to be heated to temperature $(>100\,^\circ \mathrm {C})$. A preliminary set of experiments was done with arrowroot starch (Indo Moulders, India) available commercially for stiffening fabrics.
2.5gm of the starch is dissolved in 100ml of distilled water and it is heated up for 20 minutes and boiled for 2 minutes to get a clear gel. A pinch of dye is added to enhance the contrast and the solution is allowed to cool. Continuous stirring is necessary so that lumps are not formed and a smooth homogenous fluid is obtained. 

The second set of experiments was done on a well characterized sample of potato starch
($(C_6 H_{10}O_5)_m$) manufactured by Lobachemie Pvt. Ltd. (Mumbai),  
the procedure for preparing the gel is the same above but it has to be heated for 10 minutes.

The Newtonian fluid that we use is ethylene glycol GR,(E.Merck, India). Here  also a little food colouring is added for enhancing contrast.

\subsection{Methods}

A droplet of the non-Newtonian fluid is placed on a smooth glass plate using a micro-pipette. 
The mass of the droplet  varies from 0.04 gm to 0.06 gm.

 The fluid drop is compressed by two different loading processes.
 \begin{enumerate}
 \item A load of W kg. is placed on another glass plate (identical to the bottom plate and weighing 570 gm).Then the plate and load are placed together on the drop.
 \item The upper glass plate is first placed on the droplet. Then the weight W is placed on top of the upper plate after an interval of 2 seconds.
\end{enumerate}

Each of the non-Newtonian fluids is subjected to the two loading processes described above. Ethylene glycol is compressed by the second process.
The change in area of contact between the fluid and plate is video recorded from below. The area is measured using Image Pro Plus software and plotted as a function of time.

\section{Experimental Observations}

The strain is defined as the fractional change in area per unit area at $t = 0$.

\begin{equation}
 \epsilon(t) = \frac{a(t) - a_0}{a_0}
\end{equation}
$a(t)$ and $a_0$ being respectively the area at time $t$ and at the initial time $t=0$. Time $t=0$ is taken differently for the different loading processes. 
The results for the different sets of experiments are described below.

\subsection{Set (A) - Arrowroot gel}

\begin{figure}[ht]
\begin{center}
\includegraphics[width=7.9cm, angle=0]{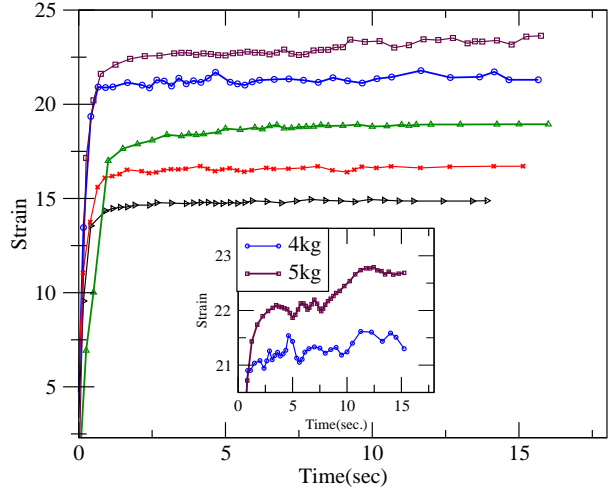}
\end{center}
\caption{\footnotesize Variation in strain with time for arrowroot gel on glass, loaded by $W$ 1 tp 5 kg by process (i) (black right triangle - 1 kg, red cross - 2 kg, green up triangle - 3 kg, blue square - 4 kg, magenta square - 5 kg) . The inset shows the erratic variations magnified for 4 and 5 kg respectively.
} 
\label{arr_tog}
\end{figure}
\begin{figure}[ht]
\begin{center}
\includegraphics[width=7.9cm, angle=0]{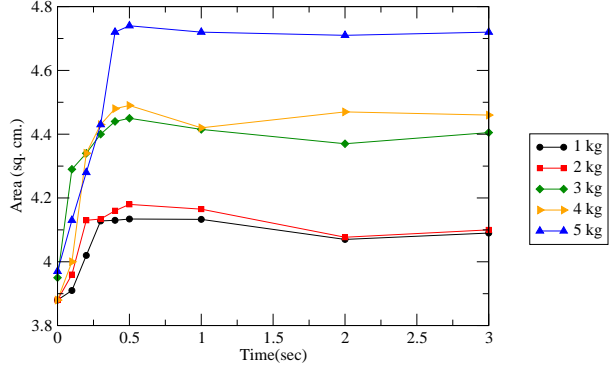}
\end{center}
\caption{\footnotesize Variation in area of contact with time for arrowroot gel on glass, loaded by $W$ 1 tp 5 kg by process (ii). Clear oscillations are seen, though not very sharp. 
} \label{arr_sep}
\end{figure}

\begin{itemize}
 
\item {Loading process (i)}
Here the initial time is the instant when the total load - (plate + W) is placed in position. 
The variation in strain for the arrowroot gel is shown in figure(\ref{arr_tog}) for different weights $W$. The area shows erratic noisy oscillations, more prominently displayed in the magnified inset. 
\item{Loading process (ii)}
Here the initial time $t=0$ refers to the instant when the weight $W$ is placed, the upper plate already being in position. Figure(\ref{arr_sep}) shows the variation in strain. Here the variation is less erratic, but all sets show that the area initially overshoots the final equilibrium value $\epsilon_\infty$, then approaches  $\epsilon_\infty$ after one or two cycles of oscillation.
\end{itemize}

This video is taken at 4 frames/sec. Though the oscillations are apparent, clearly, recording at higher speed is required. Instead of proceeding with the commercial arrowroot sample, which is not well characterized, we continued the experiments with potato starch.

\subsection{Set (B) - Potato Starch gel}
\begin{figure}[ht]
\begin{center}
\includegraphics[width=7.9cm, angle=0]{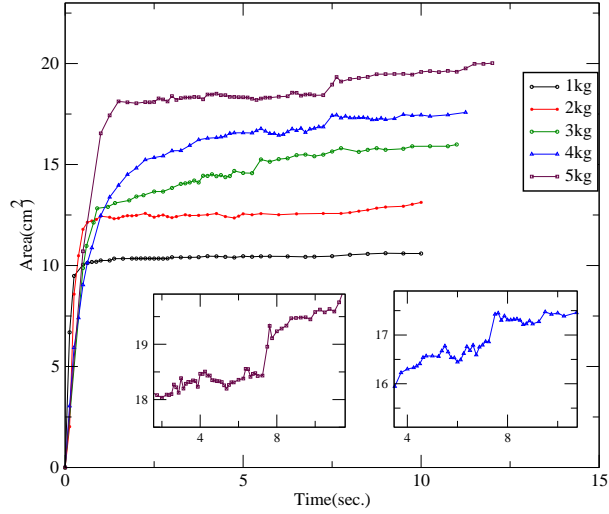}
\end{center}
\caption{\footnotesize Variation in strain with time for potato starch gel on glass, loaded by $W$ 1 tp 5 kg by process (i). The insets on the right and left show the erratic variations magnified for 4 and 5 kg respectively.
}
\label{pot_tog}
\end{figure}

\begin{figure}[ht]
\begin{center}
\includegraphics[width=7.9cm,angle=0]{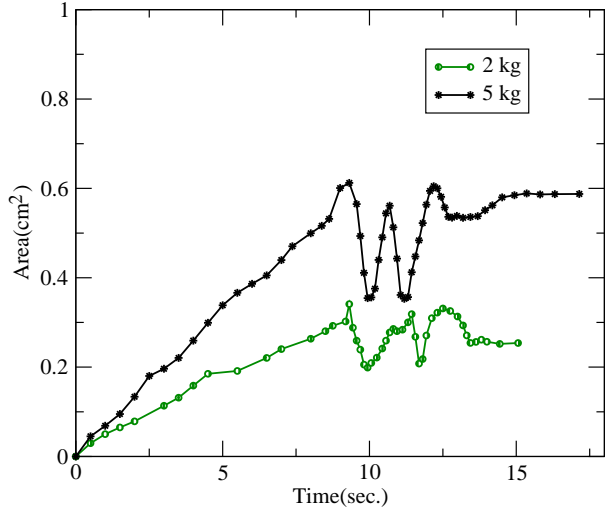}
\end{center}
\caption{\footnotesize Variation in area of contact for potato starch on glass, loaded by 2 and 5 kg by process (ii). Only results for these two values of $W$ are shown for clarity. Amplitude of the oscillations increases with load.
}
\label{all_in_one}
\end{figure}

\begin{figure}[ht]
\begin{center}
\includegraphics[width=7.9cm, angle=0]{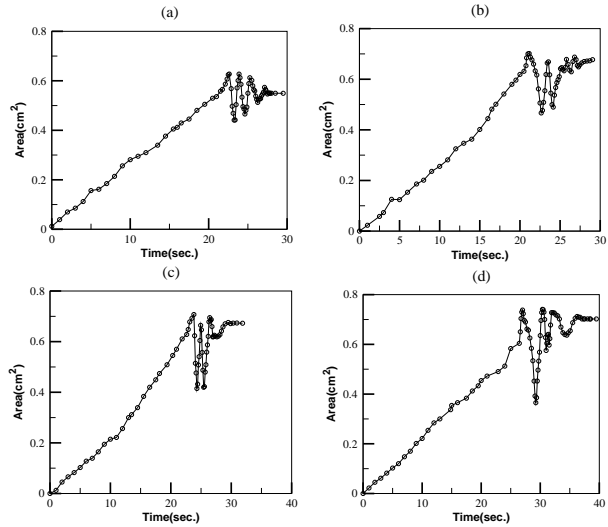}
\end{center}
\caption{Four separate experiments (a) - (d) performed identically, show the reproducibility of the nature of oscillations for potato starch gel loaded by 5 kg.}
 \label{all_5kg}
\end{figure}

All the subsequent results show videos taken at 10 frames/sec.
\begin{itemize}

\item{Loading process (i)}
Here again noisy variations, which may be superposed on oscillations are seen in the results (figure(\ref{pot_tog})).
\item{Loading process (ii)}
In this case, the erratic noise is much reduced and clear oscillations are observed in figure(\ref{all_in_one}). The oscillations die down after a few cycles to the equilibrium value $\epsilon_\infty$. The amplitude of the oscillations increases with load $W$. The experiments are repeated for reproducibility and four such data sets for $W=5$ kg are shown in figure(\ref{all_5kg}). 
\end{itemize}

\subsection{Set (C) - Ethylene glycol}
\begin{figure}[ht]
\begin{center}
\includegraphics[width=7.9cm, angle=0]{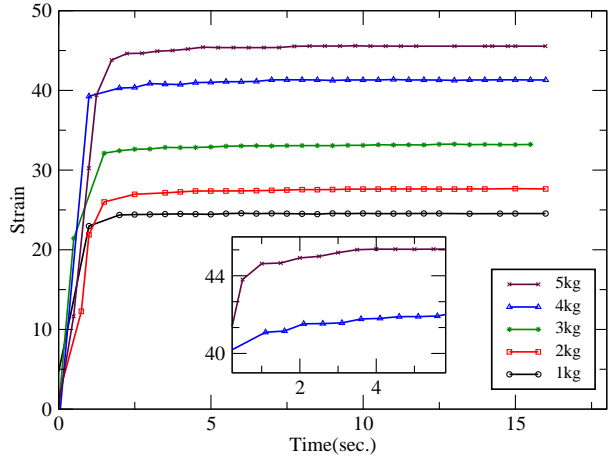}
\end{center}
\caption{Variation of strain in time for ethylene glycol loaded by different $W$ by process (ii). The inset shows that the increase is monotonic even on magnification.  }
 \label{Newtonian}
\end{figure}

Ethylene glycol has been studied earlier, together with other Newtonian fluids \cite{nag,aps}. We have repeated the experiments, recording the video at 10 frames/sec and show the results here(figure(\ref{Newtonian})) for comparison with the gels . Loading process (ii) is employed here. The results are very clearly different from the previous sets for the non-Newtonian fluids. Here, the area increases smoothly and saturates to the equilibrium value for all loads $W$.

\begin{figure}[ht]
\begin{center}
\includegraphics[width=7.9cm, angle=0]{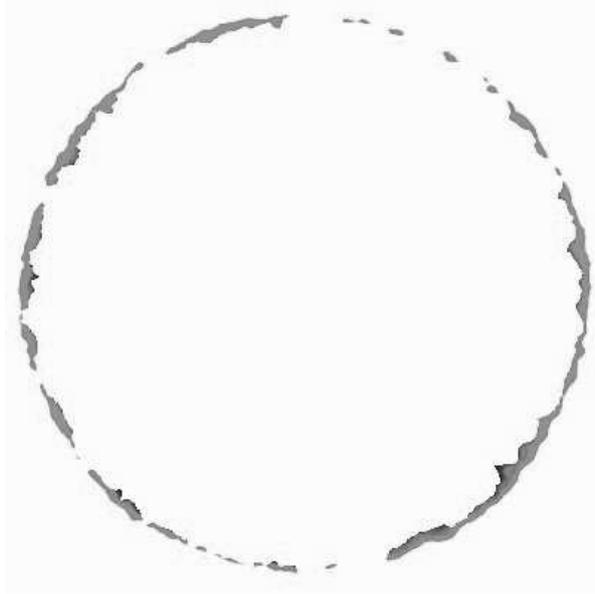}
\end{center}
\caption{\footnotesize The figure shows the difference between two outlines of the blob of potato starch gel on glass, when compressed by a 5 kg. weight. The outer contour is of a snap taken 2 seconds {\it earlier} than the inner. This shows clearly that the compressed drop shrinks before increasing again,
 }
\label{merged}
\end{figure}

The results show convincingly that these non-Newtonian fluids show an oscillatory spreading, when compressed in a Hele-Shaw cell, unlike Newtonian fluids.
It is to be noted that this is not a simple stick slip behavior, where the strain (here this is equivalent to area) would always increase, but in jumps. Here  for the gels, the strain, i.e the area, actually decreases after a short time interval during loading and again  increases.
 To establish that the film {\it does} shrink and expand in an oscillatory manner, we superpose two snapshots taken at an interval of 2 seconds, and show their difference in figure(\ref{merged}). The weight on the upper plate is 5 kg. Here the outer boundary corresponds to an instant of time {\it earlier} than the inner. So the oscillations in strain are genuine and not due to measuremental error. Further, the corrugated appearance of the boundary,  demonstrates the Saffman-Taylor like instability developed.

In the subsequent sections we suggest  a visco-elastic model to explain our experiments. At present, we focus on the less noisy data for loading process (ii). Presumably, loading process (i) needs a more complicated theory, because the weights are placed on the hemispherical drop directly. So change in shape i.e. flattening of the drop as well as change in area are to be considered in this case.

\subsection{Fluid Characterization}

It is well known that starch solutions are non-Newtonian \cite{moorthy1,singh}. It is interesting that the suspension in water made without heating is shear thickening \cite{fall}, whereas the gel made by boiling the solution is shear thinning.

Rheological studies of the arrowroot gel and potato starch gel, were done at Central Glass and Ceramic Research Institute(CGCRI, CSIR) by a Bohlin rotational Rheometer at $25 ^\circ \mathrm{C}$ temperature. 
Both show strongly shear thinning nature of the fluids with a yield stress. 
The apparent viscosity of potato starch gel is somewhat smaller than arrowroot for all strain rates.
 
\begin{figure}[ht]
\begin{center}
\includegraphics[width=7.9cm, angle=0]{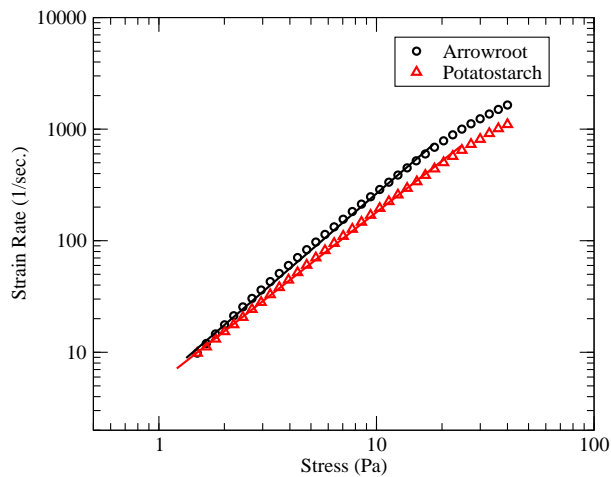}
\end{center}
\caption{\footnotesize A double logarithmic plot of strain rate versus stress for arrowroot gel and potato starch gel, show that in both cases the fluids follow a power-law over a wide range. The straight lines are best power-law fits with exponents of 1.7 and 1.5 for arrowroot and potato starch respectively.
 }
\label{log-log}
\end{figure}

 A log-log plot of strain against shear rate shows a power law relation for both fluids over a wide range (figure(\ref{log-log})). An exponent 1.7 is obtained for arrowroot, while 1.5 is obtained for potato starch, as shown in the best fit power-law straight lines  shown in figure(\ref{log-log}).
 
\section{Mathematical Modeling}

To analyze the stress-strain behavior we use a basic Kelvin-Voigt model \cite{max,bland} of a spring and dash pot in parallel as a basic unit. The viscous behavior is represented by the dash pot and the elastic nature by the spring. 
The model is further generalized by taking the $q$th derivative of the strain, thus introducing non-Newtonian rheology through fractional calculus\cite{visco,glo}. Here $q$ may be a fraction, for case  $q$ = 1 we have a Newtonian viscosity. The elastic term is assumed to be Hookean at present.
The stress is then given by

\begin{equation}
 \sigma(t) = \beta \tau^q \frac{d^q\epsilon}{dt^q} + E \epsilon
 \label{basic}
\end{equation}
Here $\epsilon$ is the strain, $E$ the elastic modulus and $\beta$ a parameter characterizing the effective viscosity of the non-Newtonian fluid and $\tau$ a characteristic time of the system.

We assume a step function to represent the loading $$ \sigma(t) = \sigma \text{ for } t\ge 0  \text{, and  }  \sigma(t) = 0 \text{ for }  t<0 $$.
The initial condition for strain is $\epsilon(t) = 0$ for $t<0 $.

The Laplace transform of equation(\ref{basic}) gives

\begin{equation}
 \epsilon(s) = \frac{\sigma}{E} \left[\frac{1}{s} - \frac{s^{q-1}}{s^q + E/\beta}\right]
 \label{lap}
\end{equation}
The inverse Laplace transform of equation(\ref{lap}) gives
\begin{equation}
 \epsilon(t/\tau) = \frac{\sigma}{E}\left[1 - ML_q\bigg(-\frac{E}{\beta} {{(t/\tau)}^q}\bigg)\right]
\label{strain}\end{equation}

where, $ML(-kt)$ is the one parameter Mittag-Leffler function defined by
$$ ML_1(-kt) = e^{-kt}$$ $$ ML_q(z) = \sum_{k=0}^\infty \frac{z^k}{\Gamma(qk+1)} $$

Visco-elastic systems typically have a `memory', the strain at time $t$ is determined by its previous loading history starting from $t = -\infty $. The Markovian system withour memory is a special case of general non-Markovian systems exhibiting correlation or anti-correlation. The general case can be treated naturally using generalized calculus methods, as follows.

One has to consider the Green's function for a general relaxation in equation(\ref{basic}), so we write the homogeneous
equation with RHS equal to zero. The strain built up for any relaxation, as a function of  a scaled dimensionless time $t_r$. 
may be treated as convolution integral of a strain variable with integral kernel  ${K_q}(t)$, as \cite{das1,das2}. We use henceforth this reduced dimensionless time $t_r$, which is the time scaled by some characteristic time of the sytem.

\begin{equation}
 \frac{d}{dt}\epsilon(\bf t_r) = -\int_0^{t_r} K_q{(t_r-t)}\epsilon(t) dt
\end{equation}

We first consider two special cases, with no memory and infinite memory.

\subsection{No memory}
If the memory kernel is $K(t_r) = B_0 \delta(t_r)$, we have the above system \ref{basic} without memory \cite{das1,das2} and the Green's function will be

$$ \epsilon(t_r) = \epsilon_0 e^{-B_0t_r} $$
that is the impulse response quickly decays to zero. Here $\epsilon_0$ is initial strain of the system at $t_r = 0$.
 
 This can be derived as follows:
 
 \begin{equation}
 K(t_r) = B_0 \delta(t_r)
\end{equation}

\begin{equation}
\frac{d}{dt}\epsilon(t_r) = -\int_0^{t_r} \delta(t-t_r)\epsilon(t) dt = -B_0\epsilon(t_r)
\end{equation}

\begin{equation}
\epsilon(t_r) = \epsilon_0 e^{-B_0t_r} 
\end{equation}

The homogeneous strain relaxation equation for no-memory case is a first order Ordinary Differential Equation i.e.

\begin{equation}
 \frac{d}{dt}\epsilon(t_r)+B_0\epsilon(t_r) = 0
\end{equation}

\subsection{Infinite memory}

If the memory kernel is a constant say $K_2(t_r) = B_2$, then we will have oscillatory Green's function, which never decays to zero .

\begin{equation}
 \frac{d^2}{dt^2}\epsilon(t) = -B_2\epsilon(t_r)
\end{equation}

\begin{equation}
 \epsilon(t_r)=\epsilon_0 cos(\sqrt{B_2}t_r)
\end{equation}

\subsection{Generalized case}
If a generalized memory integral of the following form is taken 

\begin{equation}
 K(t)=B_q t_r^{q-2} \quad ; \quad 0<q\le 2
\end{equation}
one has
\begin{equation}
 \frac{d}{dt}\epsilon(t_r)=-\frac{1}{\alpha^q}\bigg[\frac{d^{(1-q)}}{dt^{(1-q)}}\epsilon(t_r)\bigg] \label{gen}
\end{equation}
where,
\begin{equation}
 \alpha^q = \big[B_q \Gamma(q-1)\big]^{-1}
\end{equation}

Integrating the equation(\ref{gen}), we have

\begin{equation}
 \epsilon({t_r}) - \epsilon_0 = -\frac{1}{\alpha^q}\bigg[\frac{d^{(-q)}}{dt^{(-q)}}\epsilon(t_r)\bigg] 
\end{equation}

Differentiating this to order q, we get
the corresponding generalized differential equation.

 \begin{equation}
 \frac{d^q\epsilon(t)}{dt^q}-\epsilon_0\frac{t^{-q}}{\Gamma(1-q)} = -\alpha ^{-q} \epsilon(t) \label{fr} 
\end{equation}
using the fact that the differentigral of order $q$ of the constant $\epsilon_0$ gives $\epsilon_0\frac{t^{-q}}{\Gamma(1-q)} $.

This is the equation for the system with the
 memory index entering as fractional order $q$ of the Fractional Differential Equation with, $0<q\le 2$. 
 
 $q=1$ corresponds to no memory and $q=2$ corresponds to infinite memory with anti-correlation, or anti-persistence.

In equation(\ref{fr}) above, if the  initial stress be  $\epsilon_0$ , assuming Heaviside's step function as the stress input, it modifies to
equation(\ref{basic}), for $\alpha^{-q} \rightarrow B$.

The non-Newtonian fluids without oscillatory behavior will have $0<q<1$, which leads to an equation
also of fractional order, and the step-response will have monotonically increasing strain response, given by one argument
Mittag-Leffler function. Its impulse response will be a function having a long tailed decay. In other words, the response will have long-range
temporal correlation i.e. persistence.
In equation(\ref{gen}) $q$ = 0, 1 and 2 represent cases with respectively memory with  complete correlation, no memory and memory with complete anti-correlation.
These situations are analogous to a random walk with persistent memory, no memory and anti-persistent memory repectively. 

The visco-elastic system with Newtonian viscous behavior can be modeled with a discrete ideal spring and a ideal dashpot. Whereas the more complicated case with non-Newtonian viscosity
requires a different representation like a fractal chain of the ideal spring and ideal dashpot combination, or equivalently, a description using generalized calculus \cite{visco}. 
Our experiment clearly shows oscillatory
strain and thus we infer the fractional order $q$ of our system to lie between 1 and 2. Neccessity for the use of a non-trivial memory kernel clearly demonstrates that results for strain at $t$ should depend on earlier loading history, rather than the instantaneous load at $t$.

\section{Results from the Model}

\begin{figure}[ht]
\begin{center}
\includegraphics[width=7.9cm, angle=0]{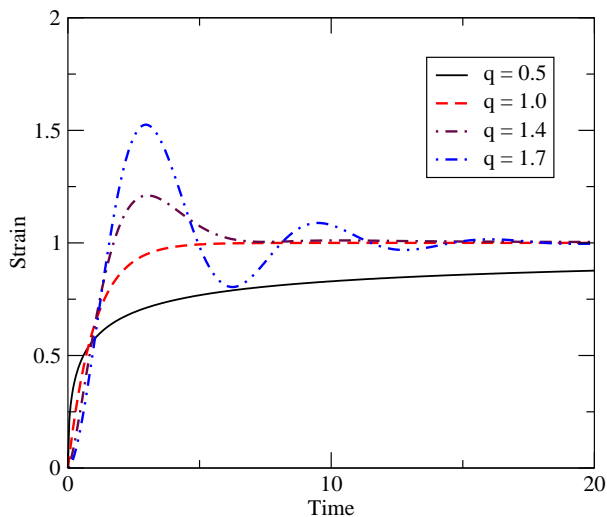}
\end{center}
\caption{\footnotesize
Variation in strain with time for different values of $q$ from equation(\ref{strain}). An oscillatory behavior is seen, which dies down to the equilibrium value.}
\label{qvar}
\end{figure}

We now plot the strain as a function of time using equation(\ref{strain}). For $q$ = 1, the strain increases smoothly and saturates to the value $\epsilon_\infty = \sigma/E$, $\sigma$ represents the load $W$ in our experiment. 
For $q < 1$, a similar behavior is observed, with a slower variation in strain. For $q > 1$ however, we see an initial increase in strain overshooting $\epsilon_\infty$ followed by oscillations before saturating to $\epsilon_\infty$. 
The oscillations are more pronounced as $q$ increases. These results are shown in figure(\ref{qvar}).
The amplitude depends on $q$ and of course the magnitude of the strain changes proportionately to the load $\sigma$, also affecting the amplitude of oscillation.

\begin{figure}[ht]
\begin{center}
\includegraphics[width=7.9cm, angle=0]{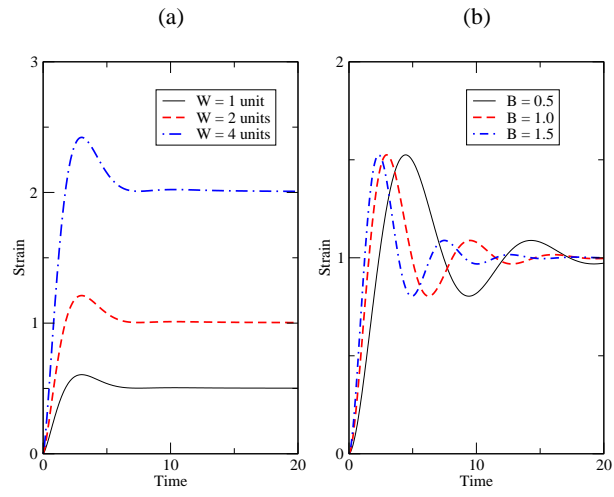}
\end{center}
\caption{\footnotesize Variation in strain with time for different values of $W$ and different values of B from equation(\ref{strain}). $q$ = 1.7 in all cases. }
\label{load_B_var}
\end{figure}

Considering the parameter  $B = \frac{E}{\beta}$ to represent the relative strengths of the elastic and viscous terms in equation(\ref{basic}), we may see how the system responds to changes in $B$. We find that variation of $B$ changes the time period of the oscillations in strain, without affecting the amplitude. Variation of strain with $W$ and $B$ are shown in figure(\ref{load_B_var}). The time period is smaller when elasticity dominates.  

If we assume that the fractional change in area is equivalent to the strain in the system, the results for the starch solutions are reproduced qualitatively by the visco-elastic model with non-Newtonian rheology. Here the qth order derivative takes care of the non-linearity in the complex fluid. In the earlier paper \cite{nag} the Newtonian fluids were assumed incompressible and the change in film thickness was calculated from the constant volume of the fluid. Here the fluid may have a finite compressibility, so we refer only to the area which is measured directly. Determination of the exact value of $q$ for our systems requires a more detailed analysis, with more information on the rheological properties. From the present study we may say $q$ lies between approximately 1.5 and 1.8.

\vskip .5cm

\section{Conclusion}
To conclude, this work demonstrates the interesting phenomenon of oscillatory  spreading of starch solutions on glass and illustrates how the approach of generalized calculus may be used to analyze it.
We plan to carry out a quantitative comparison of the theory and experiments involving the detailed rheological properties of the fluids.

\vskip 1cm

%%%%%%%%%%%%%%%%%%%%%%%%%%%%%%%%%%%%%%%%%%%%%%%%%%%%%%%%%

%%%%%%%%%%%%%%%%%%%%%
 %%%
%%%%%%%%%%%%%%%%%%%%%%%%%%%%%%%%%%%%
%%%%%%%%%%%%%%%%%%%%%%%%%%%%%%%%%%%%%%%

%%%%%%%%%%%%%%%%%%%%%%%%%%%%%%%%%%%%%%%%%%%
%%%%%%%%%%%%%%%%%%%%%%%%%%%%%%%%%%%%%%%%%%%%%%
%%%%%%%%%%%%%%%%%%%%%%%%%%%%%%%%%%%%%%%% 
%%%%%%%%%%%%%%%%%%%%%%%%%%%%%%%%%%%%%%%%%%%%%%%%%%%%%%%%%%%%%%%%%
\ack Tapati Dutta and Prof. S.P. Moulik are gratefully acknowledged for helpful discussion.
%%%%%%%%%%%%%%%%%%%%%%%%%%%%%%%%%%%%%%%%%%%%%%%%%%%%%%%%%%%%%%%%%%%%%% 
The authors thank UGC, Govt. of India for supporting this work and for providing a research grant to MDC.
%%%%%%%%%%%%%%%%%%%%%%%%%%%%%%%%%%%%%%%%%%%%

\end{document}